\documentclass[twocolumn,prd,nofootinbib,showpacs,floatfix,superscriptaddress]{revtex4}
\usepackage{epsfig}
\usepackage[dvips]{color}

\definecolor{Black}{named}{Black}
\definecolor{Red}{named}{Red}

\newcommand{\bw}{\begin{widetext}}
\newcommand{\ew}{\end{widetext}}
\begin{document}
\def\d{{\rm d}}
\def\ap{\approx}
\def\theta{\vartheta}
\def\Ecut{E_{\rm cut}}
\def\C{{\mathcal C}}

\title{Global neutrino parameter estimation using Markov Chain Monte Carlo}

\author{Steen Hannestad}
\affiliation{Department of Physics and Astronomy, University of
Aarhus, Ny Munkegade, DK--8000 Aarhus, Denmark}

\date{\today}

\begin{abstract}
We present a Markov Chain Monte Carlo global analysis of neutrino
parameters using both cosmological and experimental data. Results
are presented for the combination of all presently available data
from oscillation experiments, cosmology, and neutrinoless double
beta decay. In addition we explicitly study the interplay between
cosmological, tritium decay and neutrinoless double beta decay data
in determining the neutrino mass parameters. We furthermore discuss
how the inference of non-neutrino cosmological parameters can
benefit from future neutrino mass experiments such as the KATRIN
tritium decay experiment or neutrinoless double beta decay
experiments.
\end{abstract}


\maketitle

\section{Introduction}

The question of neutrino mass is one of the most profound in modern
particle physics. Most plausible models of neutrino mass solve the
puzzle of why neutrino masses are so small by introducing a new
scale at high energy, and precision studies of neutrino physics
therefore hold the potential to investigate physics at scales beyond
those reachable in current accelerator experiments. They also make
the study of the possible Majorana nature of neutrinos possible (see
\cite{Mohapatra:2004ht,Mohapatra:2004vr,Mohapatra:2006gs} for a
thorough discussion of this). While the neutrino mass differences
have now been measured at about 10\% precision by oscillation
experiments (see e.g.\ \cite{Maltoni:2004ei,fogli}) the absolute
mass scale remains unknown and inaccessible to oscillation
experiments.

There are, however, several possible paths to measuring the absolute
neutrino mass. The kinematical effect of neutrino mass can be probed
either via its effect on the beta decay spectrum or via its effect
on cosmological structure formation. If neutrinos are Majorana
particles a different possibility is to search for neutrinoless
double beta decay because the transition probability for this
process is proportional to the neutrino mass squared.

In the past year there have been several papers discussing how to
unify the data analysis for the various approaches
\cite{fogli,Host:2007wh}. This is a non-trivial issue, given that
completely different physics is involved and that the three probes
are actually sensitive to three distinct observables.

Here we present a new Markov Chain Monte Carlo global analysis of
neutrino parameters using both cosmological and experimental data.
The analysis software is based on the CosmoMC Markov Chain Monte
Carlo (MCMC) package for cosmological parameter estimation
\cite{Lewis:2002ah,cosmomc}, appropriately modified to incorporate
all parameters related to neutrino physics. This approach uses
Bayesian inference instead of the frequentist method commonly used
in particle physics. The approach is somewhat similar to the MCMC
technique developed in \cite{trotta} to constrain MSSM parameters.
However, a key difference is that here we keep the full cosmological
parameter estimation which allows for a closer study of the
interplay between neutrino data and cosmological parameter
estimation.

In Section II we describe the methodology used and in Section III we
present the main results for various different assumptions about
present and future data, as well as different parameter spaces.
Finally we present our conclusions in Section IV.

\section{Methodology}

The MCMC Bayesian inference approach has been described in detail
for instance in \cite{Christensen:2001gj,Lewis:2002ah}. Based on
assumed priors on each parameter it samples the likelihood function
using the Markov Chain Monte Carlo method and from that the
posterior credible intervals for all parameters can be calculated.
Before running the Markov chains it is therefore necessary to
specify both the parameters to be used and the priors on all
parameters.

For the neutrino physics part we have used the mass of the lightest
eigenstate $m_L$ ($m_L=m_1$ for the normal hierachy, $m_L=m_3$ for
the inverted hierarchy), the two mass differences $\Delta m_{12}^2$,
$\Delta m_{23}^2$ and the three mixing angles $\theta_{12}$,
$\theta_{23}$, $\theta_{13}$. We also assume that neutrinos are
Majorana particles so that there are two additional Majorana phases
$\phi_2$, $\phi_3$ \footnote{These parameters are only important
when neutrinoless double beta decay data is used}, which together
with the mass differences and mixing angles specify the observables
related to absolute neutrino mass (assuming only active neutrinos).
In total there are then 8 parameters related to the neutrino sector.

There are three separate observables related to the three different
types of probes. Cosmology is only sensitive to the neutrino mass,
and until the accuracy reaches the 0.05 eV level only to the sum of
neutrino masses (see \cite{Lesgourgues:2004ps,Lesgourgues:2006nd}
for a thorough discussion of this)
\begin{equation}
\sum m_\nu = m_1 + m_2 + m_3.
\end{equation}
In terms of the parameters used in CosmoMC this corresponds to
$m_L$, $\Delta m_{21}^2$ and $\Delta m_{32}^2$. These are therefore
the only parameters which are regarded as ``slow'' in the sense that
a change in one of them requires recalculation of the transfer
function for cosmological perturbations.

At the projected level of accuracy of KATRIN the change in the
electron energy spectrum can be described using a single effective
mass parameter which is essentially the incoherent sum (see e.g.\
\cite{Masood:2007rc})
\begin{equation}
m_\beta = \left(c_{13}^2 c_{12}^2 m_1^2 + c_{13}^2 s_{12}^2 m_2^2 +
s_{13}^2 m_3^2 \right)^{1/2}.
\end{equation}
The parameter actually measured in such experiments is in fact
$m_\beta^2$ which, being a fit-parameter, can be positive or
negative when measured.

Conversely, the effective mass measured in neutrinoless double beta
decay is the coherent sum \cite{Aalseth:2004hb,Bilenky:1987ty}
\begin{equation}
m_{\beta\beta} = \left|c_{13}^2 c_{12}^2 m_1 + c_{13}^2 s_{12}^2 m_2
e^{i \phi_2} + s_{13}^2 m_3 e^{i \phi_3} \right|,
\end{equation}
which allows for phase cancelation.

The actual parameter measured in any neutrinoless double beta decay
experiment is the half-life $T_{1/2}$ which is related to $m_{\beta
\beta}$ via the relation \cite{Rodin:2006yk}
\begin{equation}
\frac{1}{T_{1/2}} = G^{0\nu} \left|M^{0\nu}\right|^2 m_{\beta
\beta}^2,
\end{equation}
where $G^{0\nu}$ is a phase-space factor and
$\left|M^{0\nu}\right|^2$ the nuclear matrix element squared. In
principle the MCMC code should use the measured $T_{1/2}$ and the
calculated matrix element (both including uncertainties) as
parameters instead of $m_{\beta \beta}$. However, for simplicity we
assume a Gaussian error on $m_{\beta \beta}$ with the estimated
error on the matrix element from
\cite{Rodin:2006yk,Rodin:2007fz,path}.

In the following we have used cosmological parameters consistent
with the ``Vanilla'' $\Lambda$CDM model: $\Omega_b h^2$, the
physical baryon density, $\Omega_c h^2$, the physical CDM density,
$A_s$, the amplitude of primordial fluctuations, $n_s$ the scalar
spectral index, $\tau$, the optical depth to reionization, and
$H_0$, the Hubble parameter. Spatial flatness has been assumed so
that $\Omega_\Lambda = 1 - \Omega_c -\Omega_b-\Omega_\nu$. In total
there are then 6 parameters related to the cosmological model.

For the purpose of parameter estimation the publicly available
CosmoMC Markov chain monte carlo has been modified to perform
parameter estimation in this 14-dimensional parameter space. CosmoMC
has been set to use the fast/slow parameter scheme. In Table
\ref{table:priors} we give the list of parameters as well as their
priors. Note that we also include the two parameters $w$, the dark
energy equation of state, and $\alpha_s$, the running of the scalar
spectral index, both of which are discussed in Section
\ref{sec:extended}.

\begin{table}
\begin{center}
\begin{tabular}{lcll}
\hline \hline parameter & prior & & fast/slow \cr
\hline
$\log_{10}(m_L/{\rm eV})$ & -2 - 0 & Top Hat & slow \\
$\Delta m_{21}^2$ & $(4-12)\times 10^{-5}$ eV$^2$ & Top Hat & slow \\
$\Delta m_{32}^2$ & $(1-4)\times 10^{-3}$ eV$^2$ & Top Hat & slow \\
$\sin^2(\theta_{12})$ & 0-1 & Top Hat & fast \\
$\sin^2(\theta_{13})$ & 0-1 & Top Hat & fast \\
$\sin^2(\theta_{23})$ & 0-1 & Top Hat & fast \\
$\phi_2$ & 0-$2 \pi$ & Top Hat & fast \\
$\phi_3$ & 0-$2 \pi$ & Top Hat & fast \\
$\Omega_b h^2$ & 0.005-0.1 & Top Hat & slow \\
$\Omega_c h^2$ & 0.01-0.99 & Top Hat & slow \\
$\tau$ & 0.01-0.8 & Top Hat & slow \\
$n_s$ & 0.5-1.5 & Top Hat & fast \\
$\log(10^{10} A_s)$ & 2.4-4 & Top Hat & fast \\
$h_0$ & 0.3-1 & Top Hat & slow \\
$w^*$ & -2 -0 & Top Hat & slow \\
$\alpha_s^*$ & -0.2 - 0.2 & Top Hat & fast \cr \hline \hline
\end{tabular}
\end{center}
\caption{Parameters and priors used in the likelihood analysis.
Cosmological parameters marked with $^*$ are used only in Section
\ref{sec:extended}.} \label{table:priors}
\end{table}

For the observables related to neutrino oscillation data we make the
simple assumption of Gaussian errors, given by the combination of
different experiments. We note that this assumption can easily be
changed in the code and replaced with the full likelihood
calculation from experimental data.

We use the same constraints as in \cite{fogli}, given by
\begin{eqnarray}
\Delta m_{21}^2 & = & (7.92 \pm 0.71) \times 10^{-5} \, {\rm eV}^2 \nonumber \\
\Delta m_{32}^2 & = & (2.6^{+0.36}_{-0.39}) \times 10^{-3} \, {\rm eV}^2 \nonumber \\
\sin^2 \theta_{12}^2 & = & 0.314^{+0.057}_{-0.047} \label{eq:mixing}\\
\sin^2 \theta_{23}^2 & = & 0.45^{+0.16}_{-0.09} \nonumber \\
\sin^2 \theta_{13}^2 & < & 0.03, \nonumber
\end{eqnarray}
with all errors being $2\sigma$. The assumption of Gaussian errors
does not significantly alter any of our results.

\section{Results}

Based on the approach described in the previous section we have
calculated the present bound on neutrino properties using various
combinations of data sets from cosmology, tritium decay and
neutrinoless double beta decay respectively.

\subsection{Cosmological data}

Cosmological constraints on $\sum m_\nu$ have been calculated by
many different authors for various assumptions about parameters and
using different data sets (see e.g.\
\cite{Zunckel:2006mt,Cirelli:2006kt,Goobar:2006xz,Kristiansen:2006xu,%
Seljak:2006bg,Hannestad:2003xv,Hannestad:2006zg}).

Here we present just one particular example, which is exactly the
same as used in \cite{Hannestad:2007dd}. We use the WMAP CMB
temperature and polarisation data
\cite{Spergel:2006hy,Hinshaw:2006ia,Page:2006hz}, the SDSS-LRG and
2dF large scale structure data
\cite{Percival:2006gt,Tegmark:2006az,Cole:2005sx}, the SDSS-LRG
baryon acoustic oscillation data \cite{Eisenstein2005}, and the
SNI-a data set compiled in \cite{Davis:2007na}. Details about the
cosmological data can be found in \cite{Hannestad:2007dd}.

Using only the cosmological data we find a bound of $\sum m_\nu \leq
0.50$ eV for the minimal $\Lambda$CDM model. We note that this is
slightly lower than the 0.6 eV found in \cite{Hannestad:2007dd} for
the same model and data. This is worth noticing because the prior on
$m_L$ is logarithmic in the present study while it was linear in
\cite{Hannestad:2007dd}. A logarithmic prior on $m_L$ tends to
favour small $m_L$ values because of the large parameter space
volume at negative $\log_{10} m_L$ and therefore shifts the allowed
region slightly down.

This phenomenon is an integral part of Bayesian inference because a
prior probability distribution needs to be specified. In frequentist
statistics this problem does not occur and the result does not
depend on any priors. It should be noted that in the limit of
Gaussian statistics the two methods yield exactly the same result.

The phenomenon has been recently been studied in the context of
neutrino properties. For example it was shown in
\cite{Hamann:2007pi} that Bayesian inference and likelihood
maximisation give very different results for cosmological parameters
such as the radiation density as long as the likelihood function is
non-Gaussian. As more data is added and the likelihood function
approaches a Gaussian the two methods converge. The question of
Bayesian versus frequentist statistics was studied in
\cite{Host:2007wh} in the context of KATRIN. For example the
difference between a linear and a logarithmic prior on $m_\beta$ was
investigated and found to have some (not crucial) effect. In
conclusion, assumptions about priors will have an effect on the
posterior distributions as long as the likelihood function is
non-Gaussian which is the case for parameters which are not
extremely well constrained.

\begin{table*}
\begin{center}
\begin{tabular}{|l|c|c|c|c|}
\hline Parameter & Cosmo & Cosmo+HM & Cosmo+KATRIN &
Cosmo+KATRIN+GERDA \cr \hline Normal hierarchy & & & $m_{
\beta,0}=0$ & $m_{\beta\beta,0}=0$\cr \hline $\log_{\, 10} m_L$ (eV)
& $-1.334^{\, -0.795}_{\, -2.00}$ & $-1.375^{\, -0.836}_{\, -2.00}$
& $-1.376_{\, \, -2.00}^{\, \, -0.892}$ & $-1.438_{\, \, -2.00}^{\,
\, -0.959}$ \cr $\sum m_\nu$ (eV) & $0.216^{\, 0.489}_{\, 0.080}$ &
$0.205^{\, 0.447}_{\, 0.080}$ & $0.189_{\, \, 0.0802}^{\, \, 0.395}$
& $0.168_{\, \, 0.0790}^{\, \, 0.342}$ \cr $m_\beta$/$\sum m_\nu$ &
$0.271^{\, 0.331}_{\, 0.169}$ & $0.271^{\, 0.331}_{\, 0.171}$ &
$0.268_{\, \, 0.172}^{\, \, 0.330}$ & $0.260_{\, \, 0.169}^{\, \,
0.322}$ \cr $m_{\beta \beta}$ (eV) & $0.0460^{\, 0.128}_{\, 0.00}$ &
$0.0414^{\, 0.107}_{\, 0.00}$ & $0.0392_{\, \, 0.00}^{\, \, 0.101}$
& $0.0317_{\, \, 0.00}^{\, \, 0.0798}$ \cr $\Omega_c h^2$ &
$0.111^{\, 0.112}_{\, 0.103}$ & $0.111^{\, 0.117}_{\, 0.105}$ &
$0.111_{\, \, 0.104}^{\, \, 0.117}$ & $0.111_{\, \, 0.104}^{\, \,
0.117}$ \cr $n_s$ & $0.951^{\, 0.981}_{\, 0.922}$ & $0.951^{\,
0.981}_{\, 0.921}$ & $0.951_{\, \, 0.922}^{\, \, 0.980}$ &
$0.951_{\, \, 0.922}^{\, \, 0.980}$ \cr \hline Inverted hierarchy &
& & $m_{\, \beta,0}=0$ & $m_{\beta\beta,0}=0$\cr \hline $\log_{\,
10} m_L$ (eV) & $-1.35^{\, -0.792}_{\, -2.00}$ & $-1.39^{\,
-0.856}_{\, -2.00}$ & $-1.41_{\, \, -2.00}^{\, \, -0.911}$ &
$-1.47_{\, \, -2.00}^{\, \, -0.980}$ \cr $\sum m_\nu$ (eV) &
$0.232^{\, 0.501}_{\, 0.116}$ & $0.212^{\, 0.435}_{\, 0.116}$ &
$0.199_{\, \, 0.109}^{\, \, 0.419}$ & $0.182_{\, \, 0.110}^{\, \,
0.391}$ \cr $m_\beta$/$\sum m_\nu$ & $0.381^{\, 0.446}_{\, 0.338}$ &
$0.385^{\, 0.447}_{\, 0.340}$ & $0.386_{\, \, 0.340}^{\, \, 0.450}$
& $0.394_{\, \, 0.341}^{\, \, 0.451}$ \cr $m_{\beta \beta}$ (eV) &
$0.0611^{\, 0.132}_{\, 0.0230}$ & $0.0554^{\, 0.112}_{\, 0.0228}$ &
$0.0538_{\, \, 0.020}^{\, \, 0.116}$ & $0.0477_{\, \, 0.019}^{\, \,
0.101}$ \cr $\Omega_c h^2$ & $0.111^{\, 0.117}_{\, 0.104}$ &
$0.111^{\, 0.117}_{\, 0.104}$ & $0.111_{\, \, 0.104}^{\, \, 0.117}$
& $0.111_{\, \, 0.104}^{\, \, 0.117}$ \cr $n_s$ & $0.951^{\,
0.980}_{\, 0.921}$ & $0.951^{\, 0.981}_{\, 0.922}$ & $0.951_{\, \,
0.922}^{\, \, 0.981}$ & $0.951_{\, \, 0.922}^{\, \, 0.982}$ \cr
\hline Normal hierarchy & & &\multicolumn{2}{c|}{$m_{\,
\beta,0}=0.28$ eV, $m_{\beta\beta,0}=0.18$ eV}\cr \hline $\log_{\,
10} m_L$ (eV) & x & x & x & $-0.660_{\, \, -0.800}^{\, \, -0.552}$
\cr $\sum m_\nu$ (eV) & x & x & x & $0.674^{\, 0.846}_{\, 0.484}$
\cr $m_\beta$/$\sum m_\nu$ & x & x & x & $0.330_{\, \, 0.328}^{\, \,
0.332}$ \cr $m_{\beta \beta}$ (eV) & x & x & x & $0.166_{\, \,
0.0873}^{\, \, 0.244}$ \cr $\Omega_c h^2$ & x & x & x & $0.115_{\,
\, 0.108}^{\, \, 0.122}$ \cr $n_s$ & x & x & x & $0.950_{\, \,
0.921}^{\, \, 0.980}$ \cr \hline Inverted hierarchy & &
&\multicolumn{2}{c|}{$m_{\beta,0}=0.28$ eV, $m_{\beta\beta,0}=0.18$
eV} \cr \hline $\log_{\, 10} m_L$ (eV) & x & x & x & $-0.681_{\, \,
-0.857}^{\, \, -0.563}$ \cr $\sum m_\nu$ (eV) & x & x & x &
$0.651^{\, 0.829}_{\, 0.435}$ \cr $m_\beta$/$\sum m_\nu$ & x & x & x
& $0.337_{\, \, 0.335}^{\, \, 0.340}$ \cr $m_{\beta \beta}$ (eV) & x
& x & x & $0.165_{\, \, 0.0842}^{\, \, 0.243}$ \cr $\Omega_c h^2$ &
x & x & x & $0.115_{\, \, 0.108}^{\, \, 0.121}$ \cr $n_s$ & x & x &
x & $0.949_{\, \, 0.920}^{\, \, 0.980}$ \cr \hline
\end{tabular}
\end{center}
\caption{The mean value and 95\% lower and upper credible intervals
for various parameters and combinations of data.}
\label{table:parameters}
\end{table*}

\subsection{Neutrinoless double beta decay}

The upper bound on the effective neutrino mass provided by the
Heidelberg-Moscow (HM) experiment provides an additional and
comparable constraint on the absolute neutrino mass scale
\cite{KlapdorKleingrothaus:2000sn}.

We use the constraint
\begin{equation}
m_{\beta \beta} < 0.27 \, {\rm eV} \,\,\, (90\%),
\end{equation}
based on the nuclear matrix element calculation in
\cite{Rodin:2007fz}. Note that this mass range is more restrictive
than what was used in \cite{fogli} because the theoretically
predicted half-life has been corrected downwards in
\cite{Rodin:2007fz} compared to \cite{Rodin:2006yk}. We stress again
that the conversion of half-life to effective mass $m_{\beta \beta}$
depends strongly on the nuclear matrix element and that the bound
used here could turn out to be too restrictive.

As can be seen from Table \ref{table:parameters}, adding the HM data
does shift the allowed range on $\sum m_\nu$ and $m_{\beta \beta}$
down. Since the best fit cosmological model in any case has $\sum
m_\nu=0$ it has no influence on other parameters such as $\Omega_c
h^2$ and $n_s$.

Note that we have not derived any cosmological constraint based on
the claimed positive evidence from Heidelberg-Moscow
\cite{Klapdor-Kleingrothaus:2001ke,Klapdor-Kleingrothaus:2004wj,%
VolkerKlapdor-Kleingrothaus:2005qv}. Using the same assumptions as
above on the nuclear matrix element the claimed evidence translates
roughly into $0.25 \, {\rm eV} < m_{\beta \beta} < 0.5 \, {\rm eV}$
(90\% C.L.).

\begin{figure*}[htb]
\begin{center}
\epsfig{file=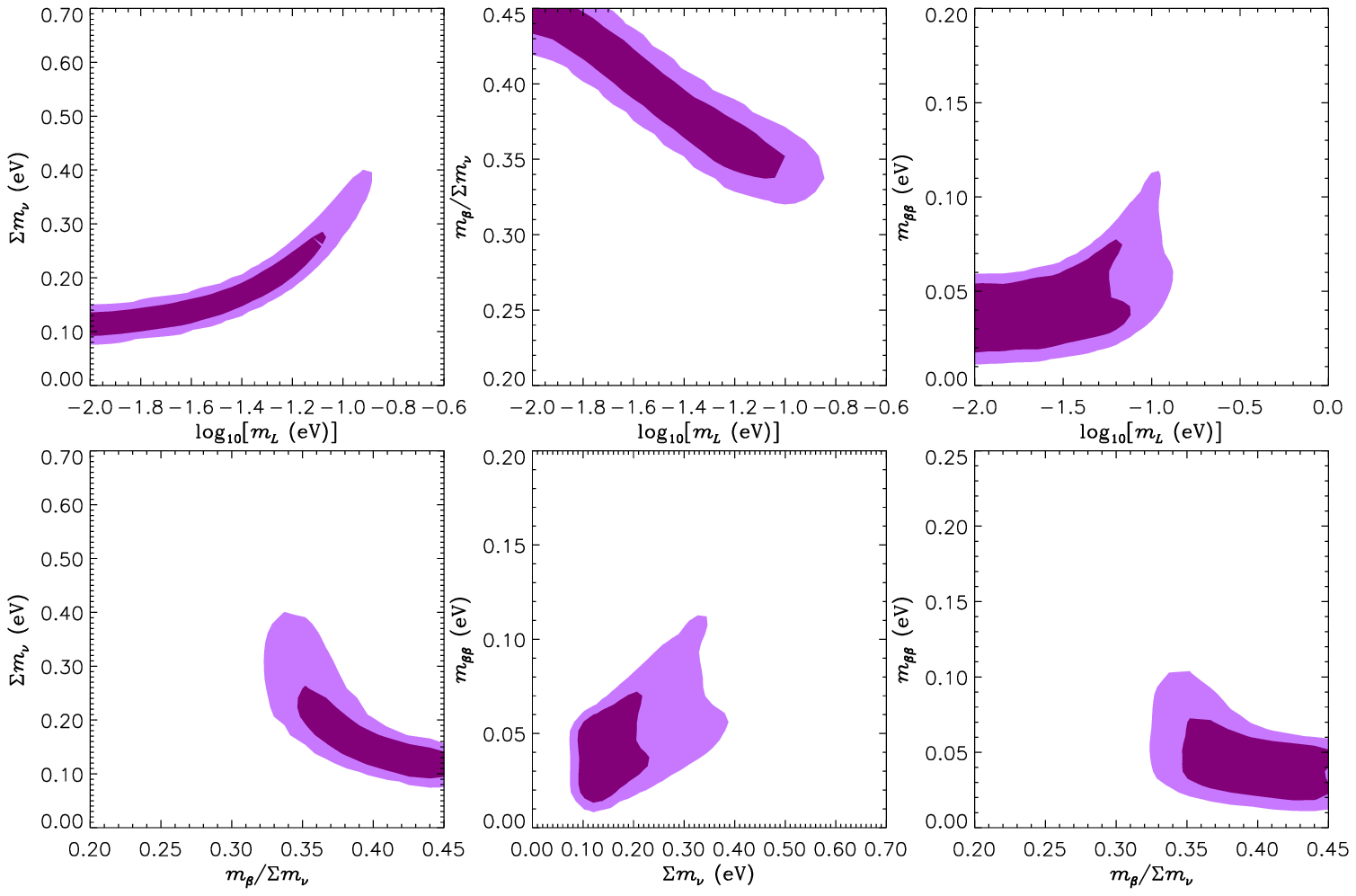,width=15.5cm}
\end{center}
\caption{68\% and 95\% contours for the Cosmo+KATRIN+GERDA data,
assuming best fit values of $m_\beta = m_{\beta \beta}=0$ (case a).
Inverted hierarchy is assumed.}\label{fig:like1}
\end{figure*}

\subsection{Future constraints - KATRIN and GERDA}

To get a better idea about the future interplay between the three
different methods for measuring the absolute mass scale we have
performed similar likelihood analyses for the presently available
cosmological data together with forecasts for the KATRIN beta decay
experiment \cite{KATRIN,Osipowicz:2001sq} and the GERDA neutrinoless
double beta decay experiment \cite{GERDA,Schonert:2005zn}.

For KATRIN we assume a Gaussian $1 \sigma$ error on $m_{\beta}^2$ of
$\sigma(m_\beta^2)=0.025 \, {\rm eV}^2$, roughly in accordance with
what was used in \cite{Host:2007wh,Kristiansen:2007di}.

For the GERDA neutrinoless double beta decay experiment we assume a
Gaussian error on $m_{\beta \beta}^2$ of 0.01 eV$^2$, corresponding
roughly to GERDA phase 2 \cite{GERDA}. We note that other
neutrinoless double beta decay experiments such as MAJORANA and
CUORE \cite{Ardito:2005ar,Gaitskell:2003zr} will reach roughly the
same sensitivity in a broadly comparable time-frame (see e.g.\
\cite{Aalseth:2004hb}).

Using these very rough experimental characteristics of the two
experiments we have proceeded to calculate constraints on neutrino
parameters using two different assumptions: \\
{\bf (a)} In this case we assume no positive detection from either
experiment so that the best fit values are $m_\beta=m_{\beta
\beta}=0$, \\
{\bf (b)} Here we assume a positive detection from both experiments,
$m_\beta^2=0.079$ eV$^2$ and $m_{\beta \beta}^2 = 0.032$ eV$^2$.
\\
The last case would for instance be realised in a model with normal
hierarchy, $m_L=m_1= x$. For both cases we perform parameter
estimation assuming both normal and inverted hierarchy.

In the first case (a) we note that KATRIN alone would significantly
tighten the cosmological constraint on $\sum m_\nu$ and GERDA would
improve this even further. $m_{\beta \beta}$ is, as expected, mainly
constrained by adding the GERDA data.

The second case (b) is more interesting from the perspective of
combining data sets. The best fit values both correspond to roughly
$2\sigma$ evidence for non-zero $m_\beta$ and $m_{\beta \beta}$
respectively. However, with the combined data $m_{\beta \beta}=0$ is
excluded at roughly $4.5 \sigma$, likewise $m_L=0$ is excluded at a
similar significance.

This exercise clearly shows the advantage of analysing all neutrino
parameters in this global way, instead of simply adding constraints.
It should be noted that since $\sum m_\nu \sim 0$ is the best fit to
present cosmological data the case (b) has a best-fit $\chi^2$ which
is higher than case (a) by $\Delta \chi^2 = 3.1$, i.e.\ it gives a
slightly (not substantially) worse fit to cosmological data.

\begin{figure*}[htb]
\begin{center}
\epsfig{file=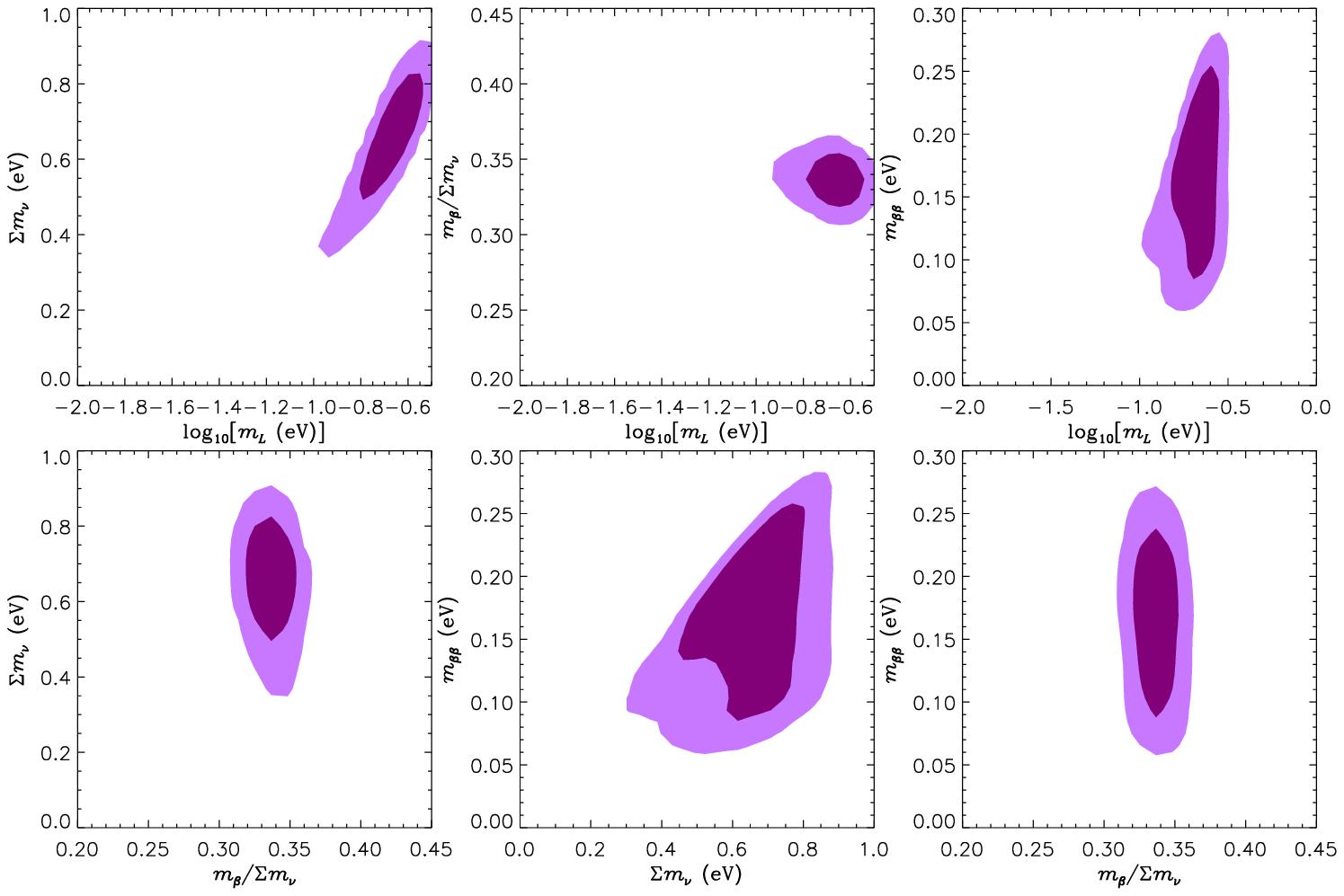,width=15.5cm}
\end{center}
\caption{68\% and 95\% contours for the Cosmo+KATRIN+GERDA data,
assuming best fit values of $m_\beta = 0.28$ eV, $m_{\beta \beta}=
0.18$ eV (case b). Inverted hierarchy is assumed.}\label{fig:like2}
\end{figure*}

In Figs.~\ref{fig:like1} and \ref{fig:like2} we show the likelihood
contours for cases (a) and (b) for the assumption of normal
hierarchy. In both cases we have used the present uncertainties on
the parameters of the mixing matrix (specified in
Eq.~\ref{eq:mixing}) which means that the non-trivial behaviour for
small $m_L,m_{\beta \beta}$ cannot be resolved. Given future
improved constraints from reactor or long baseline experiments this
region will look significantly different (see e.g.\
\cite{Pascoli:2005zb,Lindner:2005kr} for a thorough discussion of
this point).

\subsubsection{KATRIN and GERDA only}

To complete this section we have done a parameter study of cases (a)
and (b) using only KATRIN and GERDA data, excluding cosmological
constraints. The results of this can be seen in Table
\ref{table:parameter2}. For case (a) where there is no detection
from either experiment the combination of cosmological data with
KATRIN and GERDA slightly strengthens the bound on parameters, but
since they center on the same best fit value there is no marked
difference when cosmological data is added.

However, this changes completely when case (b) is studied. Here,
KATRIN and GERDA data prefer a higher value for $\sum m_\nu$ and the
combination of all three data sets significantly shift the allowed
range for all of the neutrino parameters. We note that other
cosmological parameters such as $\Omega_c h^2$ and $n_s$ are not
affected in any way when the minimal $\Lambda$CDM model is assumed.
This conclusion does not hold when larger cosmological parameter
sets are used, a point discussed in the next subsection.

\begin{table*}
\begin{center}
\begin{tabular}{|l|c|c|}
\hline Parameter & Cosmo+KATRIN+GERDA & KATRIN+GERDA \cr \hline
Normal hierarchy & \multicolumn{2}{c|}{$m_{\beta,0}=0$, $m_{
\beta\beta,0}=0$}\cr \hline $\log_{\, 10} m_L$ (eV) & $-1.438^{\,
-0.959}_{\, -2.00}$ & $-1.476^{\, -0.968}_{\, -2.00}$ \cr $\sum
m_\nu$ (eV) & $0.168^{\, 0.342}_{\, 0.079}$ & $0.157^{\, 0.335}_{\,
0.0789}$ \cr $m_\beta$/$\sum m_\nu$ & $0.260^{\, 0.322}_{\, 0.169}$
& $0.254^{\, 0.322}_{\, 0.168}$  \cr $m_{\beta \beta}$ (eV) &
$0.0317^{\, 0.0798}_{\, 0.00}$ & $0.0293^{\, 0.0757}_{\, 0.00}$ \cr
\hline Normal hierarchy & \multicolumn{2}{c|}{$m_{\beta,0}=0.28$ eV,
$m_{\beta\beta,0}=0.18$ eV}\cr \hline $\log_{\, 10} m_L$ (eV) &
$-0.660^{\, -0.552}_{\, -0.800}$ & $-0.571^{\, -0.493}_{\, -0.670}$
\cr $\sum m_\nu$ (eV) & $0.674^{\, 0.846}_{\, 0.484}$ & $0.817^{\,
0.969}_{\, 0.648}$ \cr $m_\beta$/$\sum m_\nu$ & $0.330^{\,
0.332}_{\, 0.328}$ & $0.331^{\, 0.332}_{\, 0.330}$ \cr $m_{\beta
\beta}$ (eV) & $0.166^{\, 0.244}_{\, 0.0873}$ & $0.188^{\,
0.275}_{\, 0.104}$ \cr \hline
\end{tabular}
\end{center}
\caption{The mean value and 95\% lower and upper credible intervals
for various parameters and combinations of data.}
\label{table:parameter2}
\end{table*}

\begin{table}
\begin{center}
\begin{tabular}{|l|c|}
\hline Parameter & Cosmo($+w+\alpha_s$)+KATRIN+GERDA  \cr \hline
Normal hierarchy & $m_{\beta,0}=0.28$ eV, $m_{\beta\beta,0}=0.18$ eV
\cr \hline $\log_{\, 10} m_L$ (eV) & $-0.635^{\, -0.539}_{\,
-0.779}$ \cr $\sum m_\nu$ (eV) & $0.731_{\, 0.537}^{\, 0.898}$ \cr
$m_\beta$/$\sum m_\nu$ & $0.330^{\, 0.332}_{\, 0.328}$ \cr $m_{
\beta \beta}$ (eV) & $0.176^{\, 0.252}_{\, 0.0977}$ \cr $w$ &
$-1.17^{\, -1.39}_{\, -0.99}$ \cr \hline
\end{tabular}
\end{center}
\caption{The mean value and 95\% lower and upper credible intervals
for case (b) with the larger cosmological parameter space.}
\label{table:parameter3}
\end{table}

\subsection{Extended cosmological models}
\label{sec:extended}

In order to illustrate the relation between neutrino experiments and
cosmological parameter estimation we have performed the same
analysis as before, but now adding two additional cosmological
parameters to the fit: $w$, the dark energy equation of state, and
$\alpha_s$, the running of the scalar spectral index (giving a total
of 16 parameters in the MCMC analysis). Particularly $w$ is known to
be degenerate with $\sum m_\nu$ and therefore any independent
information on $\sum m_\nu$ from experiments is potentially
important for dark energy physics. This particular degeneracy has
been studied quite extensively in recent literature. The most recent
example is \cite{Kristiansen:2007di} where the impact of a positive
KATRIN detection of $m_\beta$ on the estimation of $w$ is discussed.

\begin{figure}[h]
\begin{center}
\epsfig{file=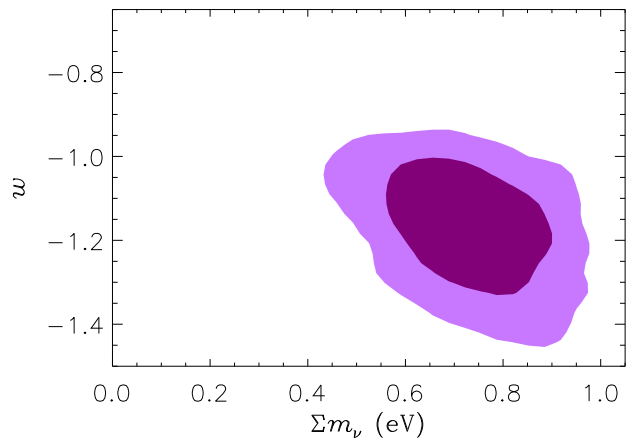,width=8.2cm}
\end{center}
\caption{68\% and 95\% contours for the
Cosmo($+w+\alpha_s$)+KATRIN+GERDA data in the $\sum m_\nu - w$
plane, assuming best fit values of $m_\beta = 0.28$ eV, $m_{\beta
\beta}= 0.18$ eV (case b). Normal hierarchy is
assumed.}\label{fig:deg1}
\end{figure}

\begin{figure}[h]
\begin{center}
\epsfig{file=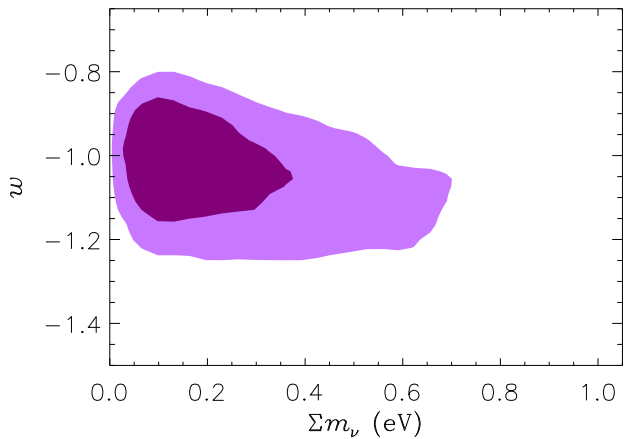,width=8.2cm}
\end{center}
\caption{68\% and 95\% contours for the Cosmo($+w+\alpha_s$) data in
the $\sum m_\nu - w$ plane. Normal hierarchy is
assumed.}\label{fig:deg2}
\end{figure}

In Fig.~\ref{fig:deg1} we show the degeneracy between $\sum m_\nu$
and $w$ for our case (b), assuming normal hierarchy. The
corresponding numbers are shown in Table \ref{table:parameter3}. The
results confirm previous findings, i.e.\ that a strongly negative
equation of state for dark energy can be compensated by increasing
the neutrino mass \cite{Hannestad:2005gj,DeLaMacorra:2006tu}. This
also means that the allowed region of $w$ for case (b) is shifted to
more negative values, in this case only marginally allowing a
cosmological constant (the 1D 95\% credible interval is $-1.39 < w <
-0.99$). The present result compares well with what is obtained in
\cite{Kristiansen:2007di}, although the assumed best fit values are
slightly different. Note also that our treatment of cosmological
data is slightly different from \cite{Kristiansen:2007di} because we
use the full BAO correlation function instead of the $A$ functional
parametrisation. For comparison, in Fig.~\ref{fig:deg2} we show the
$\sum m_\nu - w$ degeneracy for cosmological data only. In this case
we find the result $\sum m_\nu < 0.56$ eV and $-1.22 < w < -0.88$,
both at 95\% C.L., a result completely consistent with what was
found in \cite{Goobar:2006xz} for the same model and data, but using
maximisation instead of marginalisation. Note also that in this
extended model the best fit $\chi^2$ increases by 2.6 compared to
the case where only cosmological data is used (compared to 3.1 in
the smaller parameter space discussed above), i.e.\ in the extended
model the inconsistency between cosmology and the assumed positive
detection from KATRIN and GERDA is less pronounced.

The exercise carried out in this subsection clearly illustrates why
cosmological bounds on neutrino properties are model dependent. Note
that this degeneracy would be even stronger if only CMB data is
considered.

\section{Discussion}

A detailed neutrino parameter estimation study has been carried out
using the Markov Chain Monte Carlo technique with the goal of
unifying the various techniques for measuring the absolute neutrino
mass scale. The MCMC technique is extremely powerful in this regard
and allows for a very fast scanning many-dimensional likelihood
spaces. In the concrete example here we have used 8 parameters
describing the properties of light, active Majorana neutrinos, and 6
further parameters which specify the cosmology.

We find that for present data the combination of cosmological data
with the upper limit on $m_{\beta \beta}$ from Heidelberg-Moscow
slightly improves the existing cosmological bound on the sum of
neutrino masses.

More interestingly we have studied the interplay between various
future constraints from cosmology, tritium decay and neutrinoless
double beta decay. If all probes come up with a negative result the
addition of data sets does not yield any radically new information.
However, we have also studied an example in which the upcoming
KATRIN and GERDA experiments are both assumed to provide tentative
evidence for neutrino mass. In this case the combination of all
three types of data allows for a much stronger constraint on
neutrino properties than otherwise allowed.

Finally we have also studied how experimental data from tritium
decay or neutrinoless double beta decay can help in cosmological
parameter estimation, particularly concerning the dark energy
equation of state.

It should be noted that in the present analysis only presently
available cosmological data has been used. In the same time frame as
KATRIN and GERDA new cosmological data will become available and is
likely to improve the cosmological neutrino mass bound significantly
(see \cite{Xia:2007gz,Gratton:2007tb,Hannestad:2007cp,Perotto:2006rj,Takada:2005si,%
Lesgourgues:2005yv,Wang:2005vr,Song:2003gg,Hannestad:2002cn} for a
non-exhaustive list). In the somewhat longer term cosmological
constraints can be potentially be pushed below 0.1 eV sensitivity to
$\sum m_\nu$. At the same time neutrinoless double beta decay
experiments will have equally improved sensitivity and it will very
likely be possible to determine the absolute neutrino mass as well
as the nature of the mass hierarchy.

In conclusion, the combination of cosmological data with
experimental neutrino data in a global analysis will be extremely
useful in the future, when more precise experimental data becomes
available.

\section*{Acknowledgements}

Use of computing resources from the Danish Center for Scientific
Computing (DCSC) is acknowledged. Use of the CosmoMC package
\cite{Lewis:2002ah,cosmomc} is acknowledged. Amand F{\"a}ssler is
thanked for discussions on the effective neutrino mass in
neutrinoless double beta decay.



\begin{thebibliography}{99}

\bibitem{Mohapatra:2004ht}
  R.~N.~Mohapatra,
  ``Physics of neutrino mass,''
  eConf {\bf C040802}, L011 (2004)
  [New J.\ Phys.\  {\bf 6}, 82 (2004)]
  [arXiv:hep-ph/0411131].

%
\bibitem{Mohapatra:2004vr}
  R.~N.~Mohapatra {\it et al.},
  ``Theory of neutrinos,''
  arXiv:hep-ph/0412099.

\bibitem{Mohapatra:2006gs}
  R.~N.~Mohapatra and A.~Y.~Smirnov,
  ``Neutrino mass and new physics,''
  Ann.\ Rev.\ Nucl.\ Part.\ Sci.\  {\bf 56}, 569 (2006)
  [arXiv:hep-ph/0603118].

\bibitem{Maltoni:2004ei}
  M.~Maltoni, T.~Schwetz, M.~A.~Tortola and J.~W.~F.~Valle,
  ``Status of global fits to neutrino oscillations,''
  New J.\ Phys.\  {\bf 6}, 122 (2004)
  [arXiv:hep-ph/0405172].

\bibitem{fogli}
  G.~L.~Fogli {\it et al.},
  ``Observables sensitive to absolute neutrino masses: A reappraisal after
  WMAP-3y and first MINOS results,''
  Phys.\ Rev.\  D {\bf 75}, 053001 (2007)
  [arXiv:hep-ph/0608060].

%
\bibitem{Host:2007wh}
  O.~Host, O.~Lahav, F.~B.~Abdalla and K.~Eitel,
``Forecasting neutrino masses from combining KATRIN and the CMB:
Frequentist and Bayesian analyses,''
  arXiv:0709.1317 [hep-ph].


\bibitem{Lewis:2002ah}
  A.~Lewis and S.~Bridle,
  ``Cosmological parameters from CMB and other data:
  A Monte-Carlo approach,''
  Phys.\ Rev.\ D {\bf 66} (2002) 103511
  [arXiv:astro-ph/0205436]

\bibitem{cosmomc}
  A.~Lewis, Homepage, {\tt http://cosmologist.info}

\bibitem{trotta}
  R.~R.~de Austri, R.~Trotta and L.~Roszkowski,
  ``A Markov chain Monte Carlo analysis of the CMSSM,''
  JHEP {\bf 0605}, 002 (2006)
  [arXiv:hep-ph/0602028].

\bibitem{Christensen:2001gj}
  N.~Christensen, R.~Meyer, L.~Knox and B.~Luey,
  ``Bayesian Methods for Cosmological Parameter Estimation from Cosmic
  Microwave Background Measurements,''
  Class.\ Quant.\ Grav.\  {\bf 18}, 2677 (2001)
  [arXiv:astro-ph/0103134].

\bibitem{Lesgourgues:2004ps}
  J.~Lesgourgues, S.~Pastor and L.~Perotto,
  ``Probing neutrino masses with future galaxy redshift surveys,''
  Phys.\ Rev.\  D {\bf 70}, 045016 (2004)
  [arXiv:hep-ph/0403296].


\bibitem{Lesgourgues:2006nd}
  J.~Lesgourgues and S.~Pastor,
  ``Massive neutrinos and cosmology,''
  Phys.\ Rept.\  {\bf 429} (2006) 307
  [arXiv:astro-ph/0603494].

\bibitem{Masood:2007rc}
  S.~S.~Masood, S.~Nasri, J.~Schechter, M.~A.~Tortola, J.~W.~F.~Valle and C.~Weinheimer,
  ``Exact relativistic beta decay endpoint spectrum,''
  arXiv:0706.0897 [hep-ph].

%
\bibitem{Aalseth:2004hb}
  C.~Aalseth {\it et al.},
  ``Neutrinoless double beta decay and direct searches for neutrino mass,''
  arXiv:hep-ph/0412300.

\bibitem{Bilenky:1987ty}
  S.~M.~Bilenky and S.~T.~Petcov,
  ``Massive Neutrinos and Neutrino Oscillations,''
  Rev.\ Mod.\ Phys.\  {\bf 59}, 671 (1987)
  [Erratum-ibid.\  {\bf 61}, 169 (1989)].





%
\bibitem{Rodin:2006yk}
  V.~A.~Rodin, A.~Faessler, F.~Simkovic and P.~Vogel,
  ``Assessment of uncertainties in QRPA 0nu beta beta-decay nuclear matrix
  elements,''
  Nucl.\ Phys.\  A {\bf 766}, 107 (2006).

%
\bibitem{Rodin:2007fz}
  V.~A.~Rodin, A.~Faessler, F.~Simkovic and P.~Vogel,
  ``Erratum: Assessment of uncertainties in QRPA $0\nu\beta\beta$-decay nuclear
  matrix elements [Nucl. Phys. A 766, 107 (2006)],''
  Nucl.\ Phys.\  A {\bf 766}, 107 (2006)
  [arXiv:0706.4304 [nucl-th]].

\bibitem{path}A.~Faessler, talk at ''The path to neutrino mass''
workshop, Aarhus, September 2007 ({\tt
http://astroparticle.phys.au.dk})

\bibitem{Zunckel:2006mt}
  C.~Zunckel and P.~G.~Ferreira,
  ``Conservative estimates of the mass of the neutrino from cosmology,''
  arXiv:astro-ph/0610597.

\bibitem{Cirelli:2006kt}
  M.~Cirelli and A.~Strumia,
  ``Cosmology of neutrinos and extra light particles after WMAP3,''
  JCAP {\bf 0612} (2006) 013
  [arXiv:astro-ph/0607086].

\bibitem{Goobar:2006xz}
  A.~Goobar, S.~Hannestad, E.~Mortsell and H.~Tu,
  ``A new bound on the neutrino mass from the SDSS baryon acoustic peak,''
  JCAP {\bf 0606} (2006) 019
  [arXiv:astro-ph/0602155].

\bibitem{Kristiansen:2006xu}
  J.~R.~Kristiansen, H.~K.~Eriksen and O.~Elgaroy,
  ``Revised WMAP constraints on neutrino masses and other extensions of the
  minimal Lambda CDM model,''
  Phys.\ Rev.\  D {\bf 74}, 123005 (2006).

\bibitem{Seljak:2006bg}
  U.~Seljak, A.~Slosar and P.~McDonald,
  ``Cosmological parameters from combining the Lyman-alpha forest with CMB,
  galaxy clustering and SN constraints,''
  JCAP {\bf 0610}, 014 (2006)
  [arXiv:astro-ph/0604335].

\bibitem{Hannestad:2003xv}
  S.~Hannestad,
  ``Neutrino masses and the number of neutrino species from WMAP and  2dFGRS,''
  JCAP {\bf 0305}, 004 (2003)
  [arXiv:astro-ph/0303076].

\bibitem{Hannestad:2006zg}
  S.~Hannestad,
  ``Primordial neutrinos,''
  Ann.\ Rev.\ Nucl.\ Part.\ Sci.\  {\bf 56} (2006) 137
  [arXiv:hep-ph/0602058].


\bibitem{Hannestad:2007dd}
  S.~Hannestad, A.~Mirizzi, G.~G.~Raffelt and Y.~Y.~Y.~Wong,
  ``Cosmological constraints on neutrino plus axion hot dark matter,''
  JCAP {\bf 0708}, 015 (2007)
  [arXiv:0706.4198 [astro-ph]].

\bibitem{Spergel:2006hy}
  D.~N.~Spergel {\it et al.}, ``Wilkinson Microwave Anisotropy Probe
  (WMAP) three year results: Implications for cosmology,''
Astrophys.\ J.\ Suppl.\ {\bf 170} (2007) 377
  [arXiv:astro-ph/0603449].


\bibitem{Hinshaw:2006ia}
  G.~Hinshaw {\it et al.},
  ``Three-year Wilkinson Microwave Anisotropy Probe (WMAP)
  observations: Temperature analysis,''
Astrophys.\ J.\ Suppl.\ {\bf 170} (2007) 288
  [arXiv:astro-ph/0603451].

\bibitem{Page:2006hz}
  L.~Page {\it et al.},
  ``Three year Wilkinson Microwave Anisotropy Probe (WMAP)
  observations: Polarization analysis,''
Astrophys.\ J.\ Suppl.\ {\bf 170} (2007) 335
  [arXiv:astro-ph/0603450].



\bibitem{Percival:2006gt}
  W.~J.~Percival {\it et al.},
  ``The shape of the SDSS DR5 galaxy power spectrum,''
Astrophys.\ J.\ {\bf 657} (2007) 645
  [arXiv:astro-ph/0608636].

\bibitem{Tegmark:2006az}
  M.~Tegmark {\it et al.},
  ``Cosmological Constraints from the SDSS Luminous Red Galaxies,''
  Phys.\ Rev.\  D {\bf 74} (2006) 123507
  [arXiv:astro-ph/0608632].


\bibitem{Cole:2005sx}
  S.~Cole {\it et al.} [2dFGRS Collaboration],
  ``The 2dF Galaxy Redshift Survey: Power-spectrum analysis
  of the final dataset and cosmological implications,''
  Mon.\ Not.\ Roy.\ Astron.\ Soc.\  {\bf 362} (2005) 505
  [arXiv:astro-ph/0501174].


\bibitem{Eisenstein2005}
  D.~J.~Eisenstein {\it et al.} [SDSS Collaboration],
  ``Detection of the baryon acoustic peak in the
  large-scale correlation function of SDSS
  luminous red galaxies,''
  Astrophys.\ J.\  {\bf 633} (2005) 560
  [arXiv:astro-ph/0501171];
  see also
  {\tt http://cmb.as.arizona.edu/$\sim$eisenste/acousticpeak}


\bibitem{Davis:2007na}
  T.~M.~Davis {\it et al.},
  ``Scrutinizing exotic cosmological models using ESSENCE
  supernova data combined with other cosmological probes,''
  arXiv:astro-ph/0701510.

\bibitem{Hamann:2007pi}
  J.~Hamann, S.~Hannestad, G.~G.~Raffelt and Y.~Y.~Y.~Wong,
  ``Observational bounds on the cosmic radiation density,''
  JCAP {\bf 0708}, 021 (2007)
  [arXiv:0705.0440 [astro-ph]].



\bibitem{KlapdorKleingrothaus:2000sn}
  H.~V.~Klapdor-Kleingrothaus {\it et al.},
  ``Latest results from the Heidelberg-Moscow double-beta-decay experiment,''
  Eur.\ Phys.\ J.\  A {\bf 12}, 147 (2001)
  [arXiv:hep-ph/0103062].

\bibitem{Klapdor-Kleingrothaus:2001ke}
H.~V.~Klapdor-Kleingrothaus, A.~Dietz, H.~L.~Harney and
I.~V.~Krivosheina, ``Evidence for neutrinoless double beta decay,''
Mod.\ Phys.\ Lett.\ A {\bf 16}, 2409 (2001) [arXiv:hep-ph/0201231].


\bibitem{Klapdor-Kleingrothaus:2004wj}
H.~V.~Klapdor-Kleingrothaus, I.~V.~Krivosheina, A.~Dietz and
O.~Chkvorets, ``Search for neutrinoless double beta decay with
enriched Ge-76 in Gran Sasso
Phys.\ Lett.\ B {\bf 586}, 198 (2004) [arXiv:hep-ph/0404088].

\bibitem{VolkerKlapdor-Kleingrothaus:2005qv}
  H.~Volker Klapdor-Kleingrothaus,
  ``First evidence for neutrinoless double beta decay - and world status of
  double beta experiments,''
  arXiv:hep-ph/0512263.


\bibitem{KATRIN}{\tt http://www-ik.fzk.de/$\sim$katrin/index.html}

\bibitem{Osipowicz:2001sq}
  A.~Osipowicz {\it et al.}  [KATRIN Collaboration],
  ``KATRIN: A next generation tritium beta decay experiment with sub-eV
  sensitivity for the electron neutrino mass,''
  arXiv:hep-ex/0109033.

\bibitem{GERDA}{\tt http://www.mpi-hd.mpg.de/ge76/}

\bibitem{Schonert:2005zn}
  S.~Schonert {\it et al.}  [GERDA Collaboration],
  ``The GERmanium Detector Array (GERDA) for the search of neutrinoless beta
  beta decays of Ge-76 at LNGS,''
  Nucl.\ Phys.\ Proc.\ Suppl.\  {\bf 145}, 242 (2005).

\bibitem{Kristiansen:2007di}
  J.~R.~Kristiansen and O.~Elgaroy,
  ``Cosmological implications of the KATRIN experiment,''
  arXiv:0709.4152 [astro-ph].

%
\bibitem{Ardito:2005ar}
  R.~Ardito {\it et al.},
  ``CUORE: A cryogenic underground observatory for rare events,''
  arXiv:hep-ex/0501010.

\bibitem{Gaitskell:2003zr}
  R.~Gaitskell {\it et al.}  [Majorana Collaboration],
  ``White paper on the Majorana zero-neutrino double-beta decay experiment,''
  arXiv:nucl-ex/0311013.

\bibitem{Pascoli:2005zb}
  S.~Pascoli, S.~T.~Petcov and T.~Schwetz,
  ``The absolute neutrino mass scale, neutrino mass spectrum, Majorana
  CP-violation and neutrinoless double-beta decay,''
  Nucl.\ Phys.\  B {\bf 734}, 24 (2006)
  [arXiv:hep-ph/0505226].

\bibitem{Lindner:2005kr}
  M.~Lindner, A.~Merle and W.~Rodejohann,
  ``Improved limit on theta(13) and implications for neutrino masses in
  neutrino-less double beta decay and cosmology,''
  Phys.\ Rev.\  D {\bf 73}, 053005 (2006)
  [arXiv:hep-ph/0512143].

%
\bibitem{Hannestad:2005gj}
  S.~Hannestad,
  ``Neutrino masses and the dark energy equation of state: Relaxing the
  cosmological neutrino mass bound,''
  Phys.\ Rev.\ Lett.\  {\bf 95}, 221301 (2005)
  [arXiv:astro-ph/0505551].

\bibitem{DeLaMacorra:2006tu}
  A.~De La Macorra, A.~Melchiorri, P.~Serra and R.~Bean,
  ``The impact of neutrino masses on the determination of dark energy
  properties,''
  Astropart.\ Phys.\  {\bf 27}, 406 (2007)
  [arXiv:astro-ph/0608351].



%

\bibitem{Xia:2007gz}
  J.~Q.~Xia, H.~Li, G.~B.~Zhao and X.~Zhang,
  ``Probing for the Cosmological Parameters with PLANCK Measurement,''
  arXiv:0708.1111 [astro-ph].

\bibitem{Gratton:2007tb}
  S.~Gratton, A.~Lewis and G.~Efstathiou,
  ``Prospects for Constraining Neutrino Mass Using Planck and Lyman-Alpha
  Forest Data,''
  arXiv:0705.3100 [astro-ph].

\bibitem{Hannestad:2007cp}
  S.~Hannestad and Y.~Y.~Y.~Wong,
  ``Neutrino mass from future high redshift galaxy surveys: Sensitivity and
  detection threshold,''
  JCAP {\bf 0707}, 004 (2007)
  [arXiv:astro-ph/0703031].

\bibitem{Perotto:2006rj}
  L.~Perotto, J.~Lesgourgues, S.~Hannestad, H.~Tu and Y.~Y.~Y.~Wong,
  ``Probing cosmological parameters with the CMB: Forecasts from full Monte
  Carlo simulations,''
  JCAP {\bf 0610}, 013 (2006)
  [arXiv:astro-ph/0606227].

\bibitem{Takada:2005si}
  M.~Takada, E.~Komatsu and T.~Futamase,
  ``Cosmology with high-redshift galaxy survey: Neutrino mass and  inflation,''
  Phys.\ Rev.\  D {\bf 73}, 083520 (2006)
  [arXiv:astro-ph/0512374].

\bibitem{Lesgourgues:2005yv}
  J.~Lesgourgues, L.~Perotto, S.~Pastor and M.~Piat,
  ``Probing neutrino masses with CMB lensing extraction,''
  Phys.\ Rev.\  D {\bf 73}, 045021 (2006)
  [arXiv:astro-ph/0511735].

\bibitem{Wang:2005vr}
  S.~Wang, Z.~Haiman, W.~Hu, J.~Khoury and M.~May,
  ``Weighing neutrinos with galaxy cluster surveys,''
  Phys.\ Rev.\ Lett.\  {\bf 95}, 011302 (2005)
  [arXiv:astro-ph/0505390].

\bibitem{Song:2003gg}
  Y.~S.~Song and L.~Knox,
  ``Dark energy tomography,''
  arXiv:astro-ph/0312175.

\bibitem{Hannestad:2002cn}
  S.~Hannestad,
  ``Can cosmology detect hierarchical neutrino masses?,''
  Phys.\ Rev.\  D {\bf 67}, 085017 (2003)
  [arXiv:astro-ph/0211106].



\end{thebibliography}
\end{document}